\documentclass[12pt]{article}
\usepackage{amsmath,amsfonts,amssymb,graphicx,epsfig}
\newcommand{\dbar}{\kern-.1em{\raise.8ex\hbox{ -}}\kern-.6em{d}}
\newcommand{\re}{\mathop\mathrm{Re}}
\newcommand{\im}{\mathop\mathrm{Im}}

 \font\tenrm=cmr10 
\newcommand{\complex}{\kern.1em{\raise.47ex\hbox{
           $\scriptscriptstyle |$}}\kern-.40em{\rm C}}
\usepackage[usenames]{color}
\definecolor{gray}{rgb}{0.7,0.7,0.7}
\begin{document}

\title{Optimal Swimming at low Reynolds numbers}
\author{ J.E. Avron$^1$,
O. Gat$^2$,  and O. Kenneth$^1$\\
\tenrm\! ${}^1$ Department of
Physics, Technion, Haifa 32000, Israel\\
\tenrm${}^2$ Department of Electrical Engineering, Technion,
Haifa, 32000, Israel
}%
%\email{avron@physics.technion.ac.il} \twelverm
\date{\today}%
%\author{ J.E. Avron${}^1$, O. Gat${}^2$, and O. Kenneth${}^1$}
%\address{${}^1$ Department of physics, Technion, Haifa 32000, Israel
%\\ ${}^2$ Department of Electrical Engineering, Technion, Haifa
%32000, Israel}%

\maketitle %%%%%%%%%%%%%%%%%%%%%%%%%%%

\begin{abstract}
Efficient swimming at low Reynolds numbers is a major concern of
microbots. To compare the efficiencies of different swimmers we
introduce the notion of ``swimming drag coefficient'' which allows
for the ranking of swimmers. We find the optimal swimmer within a
certain class of two dimensional swimmers using conformal mappings
techniques.
\end{abstract}
%\maketitle
%\section{Introduction}
%%%%%%%%%%%%%%%%%%%%%%%%%%%%%%%%%%%%%%%%%%%%%%%%%%%%%%%%%%

%%%%%%%%%%%%%%%%%%%%%%%%%%%
{\it Motivation: }  Swimming at low Reynolds numbers is the theory
of the locomotion of small microscopic organisms
\cite{ref:lighthill,ref:purcel,ref:wilczek,ref:berg,ref:lighthil2,
ref:childress,ref:ehlers,ref:koiller,ref:shapere}. It is also
relevant to the locomotion of small robots \cite{ref:becker}.
Although microbots do not yet exist, they are part of the grand
vision of nano-science \cite{ref:feynman,ref:kosa,ref:nanomotors}
and it is important to understand the physical constraints that
underline their locomotion. Microbots must swim much faster than
bacteria if they are to interface with the macroscopic world. A
micron-size robot swimming $100$ times as fast as a bacterium, at
the modest speed of  1 mm per second, has Reynolds number $Re=\rho
UL/\mu=O(10^{-3})$, and, since power scales like $U^2$, consumes
$10^4$ more power than a bacterium. Microbots must therefore
attempt to swim as effectively as possible, and the problem we
address is how to search for effective swimming styles.

Microscopic organisms use a variety of swimming techniques: Amoeba
make large deformations of their bodies, e-coli beat flagella,
paramecia use cilliary motion, and cyanobacteria travelling
surface waves \cite{ref:berg}. One of our aims is to formulate a
criterium that can be used to compare different swimming styles
and strokes.

We show, in the context of a two dimensional model reminiscent of
Amoeba swimming, how one can find the optimal swimmer in a class
of swimmers. A movie of the optimal swimmer, can be viewed in
\cite{movie}. Our notion of optimality is closely related to a
notion of efficiency which has been extensively used in the
locomotion of microorganisms
\cite{ref:lighthill,ref:lighthil2,ref:becker} but is more general
and is applicable also to swimmers whose shape changes
substantially during the swimming stroke. It is different from a
notion of optimality introduced by Shapere and Wilczek
\cite{ref:shapere}, though the two notions become equivalent when
the amplitudes of the stroke is constrained to be small. However,
as we shall see, small strokes are never optimal.
%%%%%%%%%%%%%%%%%%%%%%%%%%%%%%%%%%%%%%%

\begin{figure}[h]
\hskip 6 cm
\includegraphics[height=8cm]{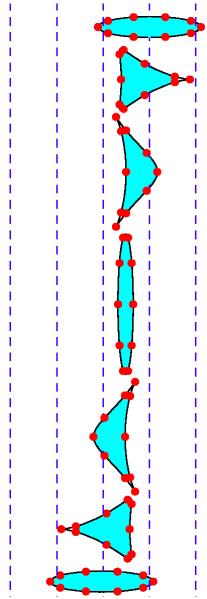}
\caption{Snapshots of the optimal swimmer shifted vertically for
visibility. The top and bottom snapshots are related by a
(horizontal) translation. The shapes with cusps correspond to
those parts of the stroke that lie on the boundary of the domain
of simple shapes. The (red) dots are fixed reference points in the
body. When the swimmer is approximately triangular, the base of
the triangle functions as an anchor that pushes or pulls the
opposite vertex. }\label{fig:optimal-swimmer}
\end{figure}

%%%%%%%%%%%%%%%%%%%%%%%%%%%%%%%%%%%%%%%%%%%%%%%%%%%%%%%%%%%%%
 {\it The theoretical framework:} Swimming results
from a periodic change of shape. We first need to recall
\cite{ref:wilczek} what is meant by a shape and a located-shape. A
located-shape is a closed surface in three dimensions (or a closed
curve in two dimensions). The surface is parameterized so each
point is marked and can be identified with a specific point of a
fixed reference, see fig.~\ref{fig:optimal-swimmer}. A shape is an
equivalence class of congruent located-shapes that differ by
translation and rotation. The space of all shapes consists of all
such equivalence classes. It is an infinite dimensional space with
a non-trivial topology. There is no a-priori metric on  the space
of shapes but, as we shall see, dissipation can be used to define
a natural metric on it.

A swimming stroke is a closed path in the space of shapes but, in
general, an open path in the space of located-shapes. We denote
the latter $\gamma(t)$, $0\le t\le\tau$. $\tau$ is the period of
the stroke. When a stroke is small the shape of the swimmer
throughout the stroke changes only a little. Once the swimmer has
completed a stroke, it is back to its original shape except that
it is translated by $X(\gamma)$ and rotated. In the problem we
consider the rotation vanishes by symmetry.

To compute the swimming step, $X(\gamma)$,  and the dissipation
$D(\gamma)$ associated to the stroke $\gamma$ one needs to solve
the (incompressible) Stokes equations for the velocity field $v$
of the ambient liquid:
\begin{equation}\label{eq:stokes}
\mu \Delta v =\nabla p, \quad \nabla \cdot v=0
\end{equation}
subject to the boundary conditions that $v$ vanishes at infinity
and satisfies a no-slip condition on the surface of the swimmer.
The no slip condition relates the liquid flow $v$ to the swimmer
movement. The latter has two parts: One comes from the rate of
change of shape and one comes from  the locomotion. Using internal forces,
the swimmer directly controls only its shape. The locomotion is determined
from the requirement that at all times the total force and torque on
the swimmer vanish \cite{ref:wilczek}.

The two dimensional case has certain special features related to
the Stokes paradox. Specifically the condition that the total
force vanishes is satisfied automatically and needs to be traded
for the condition that a regular solution of Eq.~(\ref{eq:stokes})
satisfying the boundary conditions exists, which only in two
dimensions is not automatic.
%%%%%%%%%%%%%%%%%%%%%%%%%%%

%%%%%%%%%%%%%%%%%%%%%%%%%%%%%%%%%%%%%%%%%%%%%%%%%%%%%%
{\it Optimal swimming:} Optimal swimming comes from minimizing the
energy dissipated per unit swimming distance,
$D(\gamma)/X(\gamma)$, while {\em keeping the average speed}
$X(\gamma)/\tau$ {\em fixed}. (By swimming sufficiently slowly one
can always make the dissipation arbitrarily small.) Since both the
dissipation per unit length, $D(\gamma)/X(\gamma)$, and the
velocity, $X(\gamma)/\tau$ scale as $1/\tau$ a measure for the
inefficiency of the stroke which is invariant under scaling of
$\tau$ is
\begin{equation}\label{eq:optimization}
\delta(\gamma) = \frac{\big(D(\gamma)/X(\gamma)\big)}{4\pi\mu
\big(X(\gamma)/\tau\big)}\ .
\end{equation}
The smaller $\delta(\gamma)$ the more efficient the swimmer. We
call $\delta(\gamma)$ the \emph{swimming drag coefficient}.
$\delta(\gamma)$ is a dimensionless number in two dimensions and
differs from the usual drag coefficient \cite{ref:landau} by a
factor of $Re$. In $d$ dimensions $\delta$ has dimension
$L^{d-2}$. This means that (geometrically) similar swimmers,  have
the same efficiency in two dimensions, while in three dimensions
smaller swimmers are more efficient.

$\delta$ reduces to the notion of efficiency used in the studies
of flagellar locomotion \cite{ref:lighthill} (up to numerical and
geometric factors). It is, however, different from a notion of
optimality introduced by Shapere and Wilczek \cite{ref:shapere},
where the dissipation per unit length is minimized while keeping
the {\em stroke period}, rather than the velocity, fixed.

Let $|\gamma|$ be the length of the stroke $\gamma$ in shape space measured using some metric. When a stroke is small, both $X(\gamma)$ and $D(\gamma)$ scale
like $|\gamma|^2$, independently of the choice of metric. It follows
that $\delta(\gamma)$ diverges like $1/|\gamma|^2$ for small
strokes. Small strokes are therefore inefficient. This is in
contrast with the Shapere-Wilczek criterion which  determines the
optimal stroke only up to an overall scale and does not penalize
small strokes.

%%%%%%%%%%%%%%%%%%%%%%%%%%%%%%%%%%
%%%%%%%%%%%%%%%%%%%%%%%%%%%%%%%%%%%%%%%%%%%%%5

\begin{figure}[h]
 \begin{minipage}[t]{7 cm}
 \includegraphics[width=5cm]{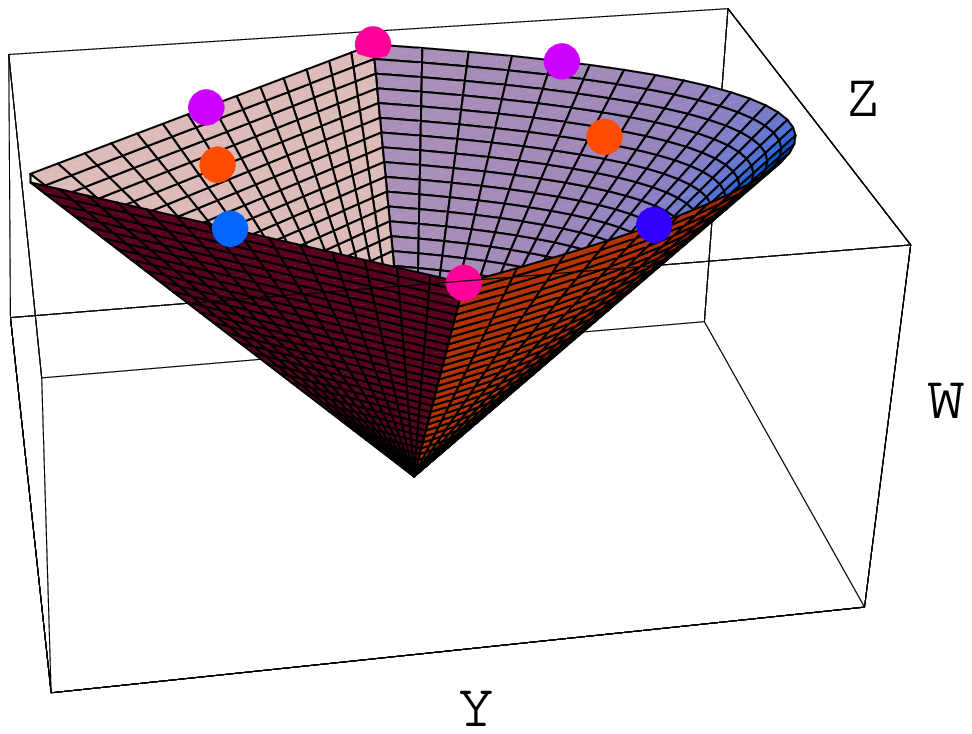}
\end{minipage}
\begin{minipage}[t]{7 cm}
\includegraphics[width=5cm]{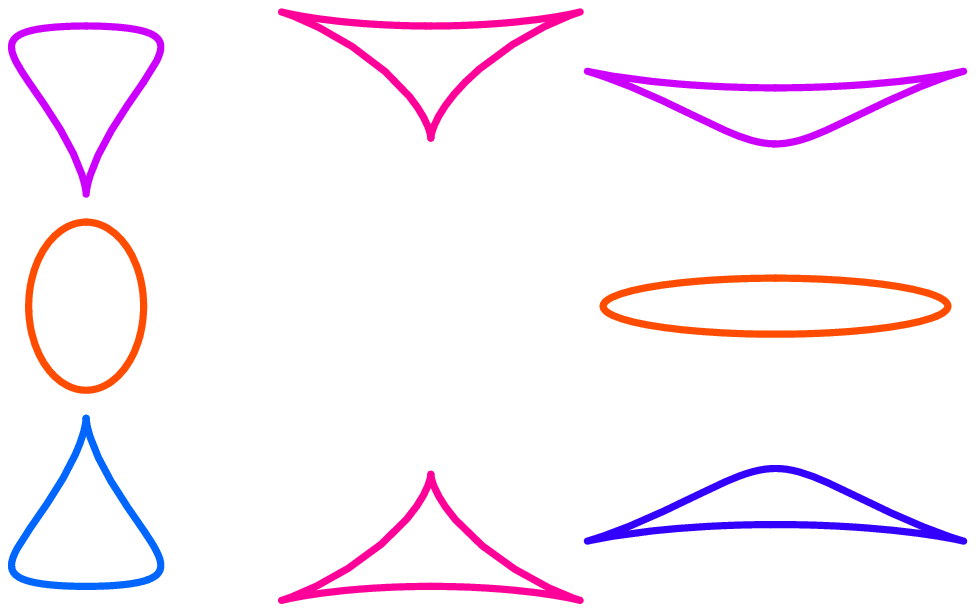}
\end{minipage}
\caption{\em  Each point $\{W,Y,Z\}$ in the cone on the left
corresponds to a two dimensional curve that does not self
intersect shown on the right.  The colors of the shapes (right)
matches the color of the dots (left). The two ellipses correspond
to the two interior points. The six shapes with cusps correspond
to the six points on the boundary of the physical
cone.}\label{fig:fish}
\end{figure}

{\it A model swimmer with a finite dimensional shape space:}
Shapere and Wilczek  \cite{ref:wilczek} introduced a class of
soluble models in two dimensions with a finite dimensional shape
space. The swimmer we consider is given by the image of the unit
disc, $|\zeta|\le1$, under a Riemann map
\begin{equation}\label{eq:z-of-zeta}
z(\zeta)= W\zeta+X+\frac{Y}\zeta+ \frac{Z}{\sqrt2\,\zeta^2}\,.
\end{equation}
As $\zeta$ traces the unit circle $z=x+iy$ traces the boundary of the (located)
shape in the complex plane (see Fig.~\ref{fig:fish} for examples). The space of
located-shapes is four dimensional with coordinates $\{W,X,Y,Z\}$.
$X$ is naturally interpreted as the position of the swimmer, since
$\{W,X,Y,Z\}$ and $\{W,0,Y,Z\}$ describe congruent curves that
differ by translation by $X$. Similarly, $\{W, Y,Z\}$ and
$e^{i\phi} \{W, Y,Z\}$, $\phi\in\mathbb{R}$, describe congruent
closed curves that differ by a rotation by $\phi$. The shape space
of the model is a space of three complex parameters defined up to
a global phase.

When $Z=0$ and $|W|\neq |Y|$ the shape is an ellipse. (When
$|W|=|Y|$ the ellipse degenerates to an interval.) A symmetry
argument shows that an elliptic swimmer can turn but can not swim.
In this sense the model with $Z\neq 0$ is a minimal model of a
swimmer.

For the sake of simplicity we shall, from now on, restrict
$\{W,Y,Z\}$ to be real. The shapes in this space are symmetric
under mirror reflection. (This follows from $\bar
z(\zeta;t)-X=z(\bar\zeta;t)-X$  where $\bar z$ denote the complex
conjugate of $z$.) A swimmer that maintains its reflection
symmetry during the stroke can not turn and can only swim in the
$x$ direction. Hence, without loss, $X(t)$ may be taken to be real
as well, and the space of shapes of Eq. (\ref{eq:z-of-zeta}) with
\emph{real} parameters, can then be identified with the three
dimensional Euclidean space $\{W,Y,Z\}\in\mathbb{R}^3$.

{\it The solution of the model:}  The stroke $\gamma$ is a
(parameterized) closed path in $\mathbb{R}^3$, i.e.
$\gamma=\big\{\big(W(t),Y(t),Z(t)\big)|0\le t\le\tau\big\}$. It
generates a  flow in the fluid surrounding the swimmer, which fills the domain
corresponding to $|\zeta|\ge 1$. The solution of the Stokes
equations, Eq.~(\ref{eq:stokes}), can then be obtained by
conformal mapping methods \cite{ref:wilczek,ref:muskhelishvili}:
\begin{equation}\label{v}
v=f_1(\zeta)+\overline{f_2(\zeta)}-z(\zeta)\,
\overline{\left(\frac{f_1^\prime(\zeta)}{z'(\zeta)}\right)},\quad
v=v_x+iv_y\,,
\end{equation}
with
\begin{equation}\label{eq:solution-1}
f_1=\frac{\dot Y}{\zeta}+\frac {\dot Z}{\sqrt 2\zeta^2}\,;
\end{equation}
\begin{equation}\label{eq:solution-2}
f_2=\dot{{X}}- \frac{\frac{W}\zeta+X+Y\zeta+
\frac{{Z}\zeta^2}{\sqrt 2}}{W-\frac Y{\zeta^2}- \frac{\sqrt 2
\,Z}{\zeta^3}} \left(\frac{\dot
Y}{\zeta^2}+\frac{\dot{Z}\sqrt{2}}{\zeta^3}\right)
+\frac{\dot{W}}\zeta\,,
\end{equation}
where dot denotes time derivative. The flow vanishes at infinity
provided $f_2(\infty)=f_1(\infty)=0$.  From
Eq.~(\ref{eq:solution-2}) one finds that this is the case
provided:
\begin{equation}\label{eq:connection}
\dbar X=A\, dY,\qquad A=\frac{Z}{\sqrt 2 W}\ .
\end{equation} This is the basic relation
between the swimming (the response), $\dbar X$, and the change in
shape (the controls), $\{W,Y,Z\}$. The notation $\dbar X$ stresses
that $X$ does not integrate to a function of $\{W,Y,Z\}$.
Geometrically, this relation is interpreted as a connection on the
space of shapes \cite{ref:wilczek}. Note that $A(W,Y,Z)$ is a
homogeneous function of degree zero.

The power $P$  dissipated by the swimmer is calculated by
integrating the stress times the velocity on the surface of the
swimmer:
\begin{equation}
P =\im\oint\bar v\left(\mu\left(\frac{\partial v}{\partial \bar
z}\right)d\bar z-pdz\right)\ ,
\end{equation}
where $p$ is the fluid pressure, $p=-4\mu\re f_1'(z)$. Using the
explicit solution given by
Eqs.~(\ref{v},\ref{eq:solution-1},\ref{eq:solution-2}), one
obtains
\begin{equation}\label{eq:dissipation}
    P=4\pi\mu\,\big(\dot W^2+\dot Y^2+\dot Z^2\big)\,.
\end{equation}
The dissipation of a stroke is then
\begin{equation} \label{eq:dgamma}
D(\gamma)=4\pi\mu\,\int_0^\tau\big(\dot W^2+\dot Y^2+\dot
Z^2\big)dt\ .
\end{equation}
%%%%%%%%%%%%%%%%%%%%%%%%
%%%%%%%%%%%%%

%%%%%%%%%%%%%%%%%%%%%%
%\vskip 1 cm

 {\it The Physical cone:}
%%%%%%%%%%%%%%%%%%%%%%%%%
A physical shape does not self-intersect. Since there are points
$\{W,Y,Z\}\in\mathbb{R}^3$ which represent curves that
self-intersect, {e.g. $Z=0$, $W=\pm Y$}, we need to remove them.
The physical shapes make a cone, for if $\{W,Y,Z\}$ does not
self-intersect neither does $\lambda \{W,Y,Z\}$ with $\lambda>0$.
The shapes associated with points in the interior of the cone are
smooth. Points on its boundary correspond to shapes with cusp-like
singularities (precursors of self-intersections). The boundary of
the physical cone is given by those $\{W,Y,Z\}$ for which there
exists $\zeta$ of unit modulus such that $z'(\zeta)=0$. One finds
that the physical cone around the $W$ axis is bounded by two
planes and a quadratic surface and is given by , see
Fig.~\ref{fig:fish},
\begin{equation}\label{eq:physicalcone}
 W\geq Y\pm \sqrt 2 \,Z\,,\quad WY\geq 2Z^2-W^2\ .
\end{equation}

%%%%%%%%%%%%%%%%%%%%%%%%%%%%%%%%%%%%%%%%%%%%%%%

{\it Optimization and orbits in magnetic fields:}
%%%%%%%%%%%%%%%%%%%%%%%%%%%%%%%%%%%%%%%%%%%%%%%%%%%%%
%\section{Large strokes are inefficient}
Admissible strokes are closed paths $\gamma$ that lie in the
physical cone. Consider the problem of minimizing the dissipation,
$D(\gamma)$, of Eq.~(\ref{eq:dissipation}) subject to the
constraint that the step size is
\begin{equation}\label{eq:xofgamma}
X(\gamma)=\oint_\gamma A \,dY\, .
\end{equation}
This is a standard problem in variational calculus. Note that
since one may set $\tau=1$ without loss, fixing the average speed
is equivalent to fixing the step size $X(\gamma)$. The minimizer,
$\gamma(t)$, must then either follow the boundary, or solve the
Euler-Lagrange equation of the functional
\begin{equation}\label{eq:action}
S_q(\gamma)=4\pi\mu\,\int_0^\tau \big(\dot W^2+\dot Y^2+\dot
Z^2\big) dt +q \int_0^\tau A\, \dot Y \, dt,
\end{equation}
where $q$ is a Lagrange multiplier. $S_q$ can be interpreted as
the action of a non-relativistic particle whose mass is $8\pi \mu$
and whose charge is $q$ moving in three dimensions under the
action of a magnetic field with vector potential $A\hat Y$.
$X(\gamma)$ may be interpreted as the flux of
$B(W,Y,Z)=\nabla\times(A\hat Y) $ through the closed path
$\gamma$.

The parameterized stroke $\gamma(t)$ that minimizes $S_q$ has
constant velocity $|\dot\gamma|=const$. (This  follows from the
fact that the flux $X(\gamma)$ is independent of parametrization
and that the action of a free particle along a one dimensional
curve is minimized at constant speed.) The dissipation is then
$D(\gamma)=4\pi\mu\frac{|\gamma|^2}\tau$, where $|\gamma|$ is the
length of the orbit, and the drag $\delta(\gamma)$ is simply:
\begin{equation}\label{eq:lgamma}
\delta(\gamma)=\left(\frac {|\gamma|} {X(\gamma)}\right)^2\ .
\end{equation}

Since the variational problem is for a domain with non-smooth
boundaries, Fig.~\ref{fig:fish}, one may worry if the minimizer
fails to be smooth. This is not the case. For if $\gamma(t)$
has a corner, smoothing the corner on a scale $\varepsilon$
shortens the length $|\gamma|$ by $O(\varepsilon)$ while
$|X(\gamma)|$ varies by $O(\varepsilon^2)$. In particular, it
follows that the minimizer avoids the corners of the physical
cone. Moreover, whenever it hits or leaves the boundary of the
physical cone, it does that tangentially, without corners.

\begin{figure}[h]
\hskip 4 cm
\includegraphics[width=6cm]{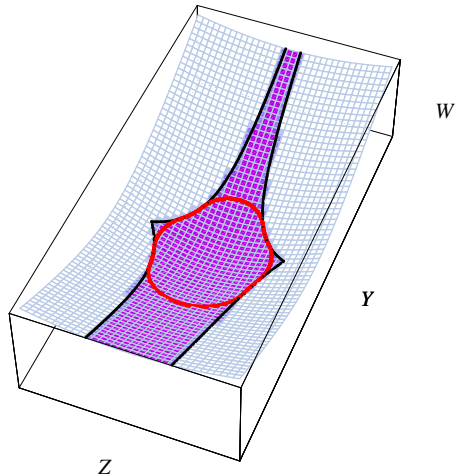}
\caption{\em The Big-Ben shaped surface is the intersection of the
physical cone (no self-intersection) with the hyperboloid of
shapes with fixed area. The boundary of the domain of
incompressible non-self-intersecting shapes is drawn black. The
inscribed red curve is the optimal stroke as computed numerically.
}\label{fig:non-compact}
\end{figure}

{\it Incompressible swimmers:}  Incompressible swimmers make a
natural (and biologically important) class of swimmers. In two
dimensions incompressibility implies constant area. The area
of the swimmer whose shape is given by Eq.~(\ref{eq:z-of-zeta}) is
\begin{equation}\label{eq:area}
\frac 1 2 \, \im\,\oint \bar z \, dz=\pi (W^2-Y^2-Z^2)\,.
\end{equation}
Fixing the area of the swimmer corresponds to restricting the
stroke to a hyperboloid in shape space. We choose the unit of area
so that the area of the swimmer is $\pi$.
The intersection of the constant area hyperboloid with the cone of
physical shapes is the Big-Ben shaped surface shown in
Fig.~\ref{fig:non-compact}. Physical strokes are represented by
closed paths that lie inside this domain, and our aim is to find
the stroke that minimizes the swimming drag $\delta(\gamma)$.

{\it Large strokes are inefficient:} The model admits strokes that
extend to arbitrarily large values of $Y$ and $W$. Since, by
Eqs.~(\ref{eq:connection},\ref{eq:physicalcone},\ref{eq:xofgamma}),
the total flux of $B$ through the Big-Ben shaped region of
Fig.~\ref{fig:non-compact} is infinite, the swimmer can swim
arbitrarily large distances with a single stroke. However, as we
presently show, large strokes are inefficient. The domain of
physical incompressible shapes for a swimmer of area $\pi$ is
contained in the strip $|Z|\le 1$. Since $A=O(\frac 1 Y)$, for $Y$
large, a long excursion, of order $\ell$, in the $Y$ direction
contributes $O(\log \ell)$ to $X(\gamma)$, but $O(\ell )$ to
$|\gamma|$. Therefore, as $\ell\to \infty$ the drag coefficient
$\delta$ diverges like $\ell/\log\ell$.

%%%%%%%%%%%%%%%%%%%%%%
%%%%%%%%%%%%%%%%%%%%%%%%%%%%%%%%%%%%%%%%%%

%%%%%%%%%%%%%%%%%%%%%%%%

{\it The minimizer:}  Since the drag diverges for small strokes
and also for large strokes (for an incompressible physical
swimmer) the minimizer of $\delta$ is a finite stroke. It can be
computed numerically using the following procedure:  Since the
minimizer is independent of the period $\tau$, one may, without
loss, restrict oneself to orbits with fixed energy, say $E=1$.
Pick the charge $q$ and find a smooth and closed orbit on the
hyperboloid of constant area with (four) sections  in  the
interior of the cone of physical shapes and (four) sections on its
boundary. There is a unique such orbit $\gamma_q$ for all $q$ that
are small enough. For each such orbit one computes, numerically,
$X(\gamma_q)$, and $|\gamma_q|$ to get $\delta(\gamma_q)$ from
Eq.~(\ref{eq:lgamma}). What remains is a minimization problem in
one variable, $q$, which yields the optimal stroke. The optimal
stroke in shape space is shown in Fig.~\ref{fig:non-compact} while
snapshots of the corresponding swimming motion in real space are
shown in Fig.~\ref{fig:optimal-swimmer}. For the optimal stroke we
find $\delta_{\rm{optimal}}\approx 9.12\ .$ For the sake of
comparison with a squirming circle consider $\gamma$ which is a
small circle of radius $r$ in the $Y-Z$ plane, centered at
$Y=Z=0,W=W_0$. One readily finds from
Eqs.~(\ref{eq:connection},\ref{eq:lgamma}) that $\delta= 8
(W_0/r)^2$. A squirmer that changes it shape by 10\% has,
$r/W_0=0.1$, and $\delta =800$.

{\it Perspective:} The optimal swimmer we have found is optimal
within the class of Riemann maps of Eq.~(\ref{eq:z-of-zeta})
satisfying incompressibility.
Enlarging the class of Riemann maps, would allow for better
swimmers. It is conceivable that there are superior swimmers that
use quite different swimming styles.  The importance of the model
lies in that it demonstrates a scheme for a systematic search of
efficient swimmers, and provides benchmark for $\delta$ for better
swimmers to beat.

%%%%%%%%%%%%%%%%%%%%%%%%%%%%%%%%%%%%%%%%%%%%%%%%%%%%%%%%%%%%%%

%%%%%%%%%%%%%%%%%%%%%%%%%%%%%%%%%%%%%%%%
{\it Acknowledgment:} We thank  E. Braun , G. Kosa, and D. Weihs
for useful discussions, E. Yariv for pointing out ref.
\cite{ref:becker} and U. Sivan for proposing the problem. This
work is supported in part by the EU grant HPRN-CT-2002-00277.

\end{document}